\documentclass[journal=amlccd,manuscript=letter]{achemso}
\setkeys{acs}{articletitle=true}

\usepackage[T1]{fontenc}       

\usepackage{graphicx}
\usepackage{dcolumn}
\usepackage{bm}
\usepackage{amsmath}
\usepackage{amsfonts}
\usepackage{amssymb}
\usepackage{braket}
\usepackage[separate-uncertainty = true, multi-part-units=single, range-phrase=--, range-units=single]{siunitx}
\usepackage{subcaption}
\usepackage{color}
\usepackage{hyperref}

\let\oldsqrt\sqrt
\def\sqrt{\mathpalette\DHLhksqrt}
\def\DHLhksqrt#1#2{%
\setbox0=\hbox{$#1\oldsqrt{#2\,}$}\dimen0=\ht0
\advance\dimen0-0.2\ht0
\setbox2=\hbox{\vrule height\ht0 depth -\dimen0}%
{\box0\lower0.4pt\box2}}

\title{Giant fluctuations in sheared viscoelastic fluids emerging in a mesoscale simulation}

\author{Airidas Korolkovas}
\affiliation{Institut Laue-Langevin, 71 rue des Martyrs, 38000 Grenoble, France}
\alsoaffiliation{Department for Physics and Astronomy, Lägerhyddsvägen 1, 752 37 Uppsala, Sweden}
\email{korolkovas@ill.fr}

\begin{document}
\begin{tocentry}
\mbox{
	\sbox0{\includegraphics{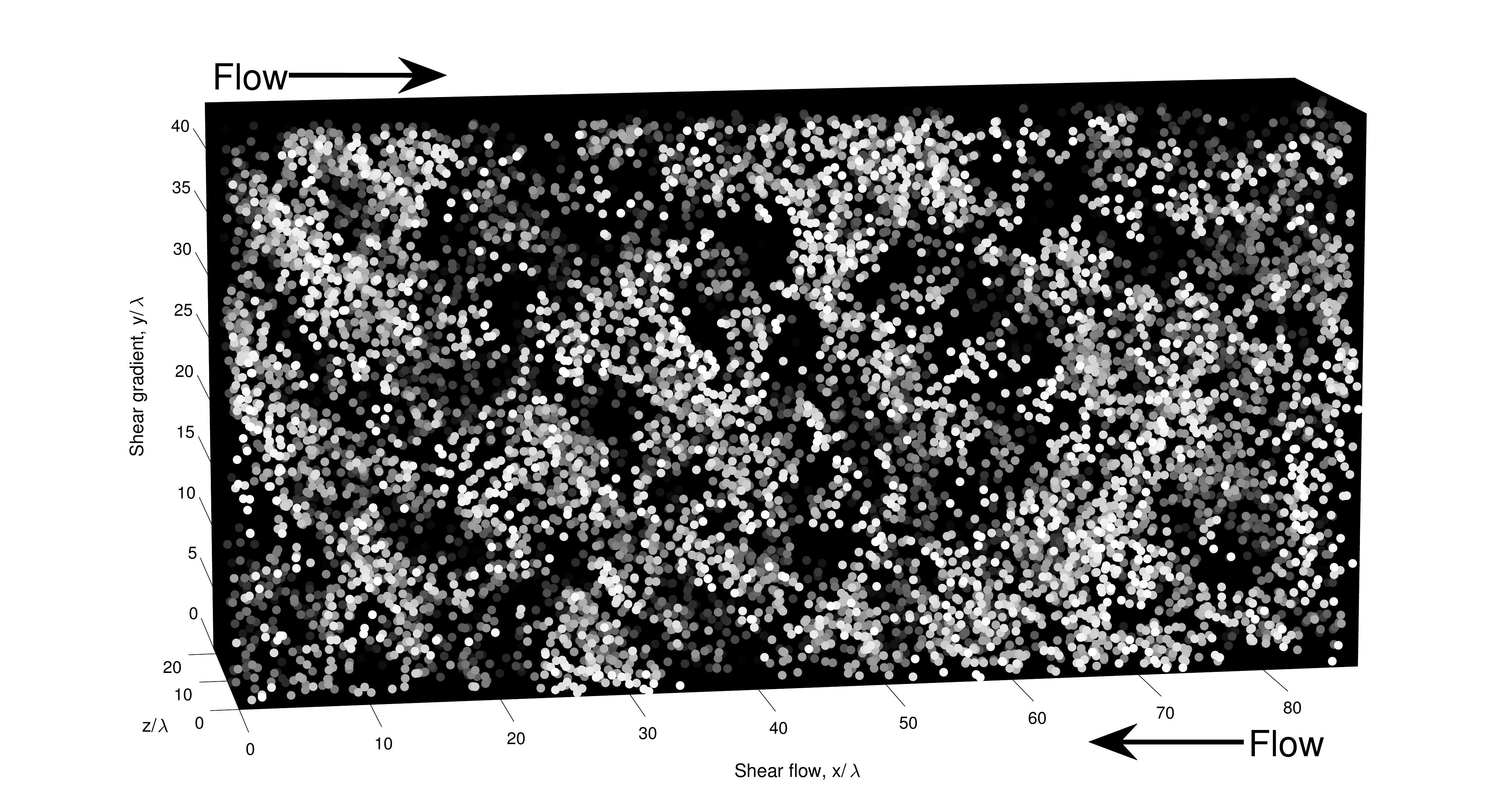}} 
\includegraphics[clip,trim={.05\wd0} 0 {.1\wd0} {.04\wd0},width=0.9\linewidth]{./fig/box.pdf}
}
\end{tocentry}



\date{\today}

\begin{abstract}
Shear flow is known to induce huge density fluctuations in otherwise clear and uniform polymer solutions. This effect is rooted in the elasticity of the entangled polymer network, and can span distances over a thousand chains wide. It has been observed in many scattering experiments, and later explained by mathematical theories. Here we inspect this phenomenon from a direct particle simulation viewpoint. The main novelty is a velocity dependent friction force, coupling the entire system and solved efficiently with sparse matrix algebra. Our minimalist model runs on a desktop PC and the results agree well with experiments.
\end{abstract}

\maketitle

\section{Introduction}
\begin{figure}[tbh!]
		\includegraphics[width=\linewidth]{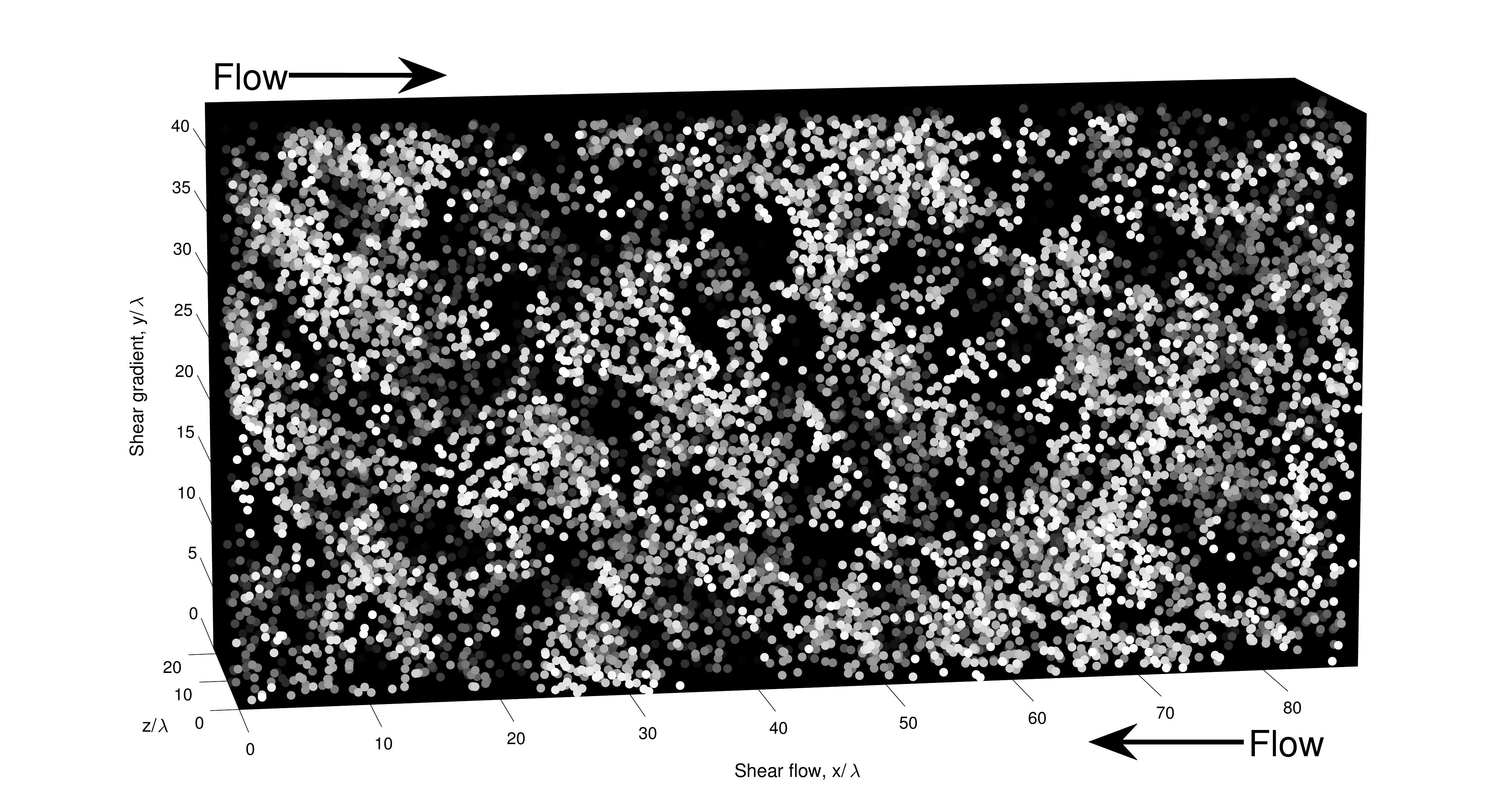}
\caption{A real-space snapshot from the simulation at a strong shear of $\kappa \tau = 0.5$, applied in the $xy$ plane. Particle brightness decreases with the coordinate $z$, to visualize depth. Notice how the clusters tend to slant backwards with respect to the shear flow, a counter-intuitive effect, well-known in entangled polymer solutions. This image may be likened to the experimental micrograph Fig.~5 in Ref.~\cite{mhetar1998slip}.}\label{box}
\end{figure}

A sugar cube in a cup of tea dissolves faster by stirring with a spoon. In contrast, certain non-Newtonian fluids defy this common sense and become less homogeneous under shear. It is evidenced by a butterfly-shaped scattering pattern of light (SALS)~\cite{hashimoto1991shear} and neutrons (SANS)~\cite{boue1994semi} emerging under a strong deformation. Examples include clay~\cite{pignon1997butterfly, schmidt2000shear, schmidt2002small, shibayama2005small}, nanoparticle~\cite{min2014microstructure} and colloidal gels~\cite{rueb1997viscoelastic}, slide ring gel~\cite{karino2005sans}, randomly cross-linked gel~\cite{horkay2000neutron}, silica-polymer~\cite{degroot1994flow}, nanoplatelet-polymer networks~\cite{lin2003shear}, carbon black filled polymers~\cite{ehrburger2001anisotropic}, and micellar worms~\cite{croce2005giant}. The biggest inhomogeneities are found in fluids with a high amount of elasticity and low osmotic pressure. As explained theoretically by Onuki~\cite{onuki1992scattering}, such fluids under shear or extensional flow develop density fluctuations on the scale much larger than the typical size of the constituent molecules.

Perhaps the simplest and the most commonly studied example is entangled polymer solutions, which are composed of flexible highly interpenetrated chains. In equilibrium, the chains spread out as uniformly as the excluded volume will allow them, and adopt the conformation of a random walk, which is the most disordered state possible (maximum entropy). The fluctuations around this equilibrium structure have been investigated with dynamic light scattering (DLS), and were found to contain two distinct phenomena: the fast and the slow modes~\cite{li2010slow, brown1990static}. The fast mode, also called cooperative diffusion (see Chapter 5 in Ref.~\cite{doi1988theory}), is the fluctuation between neighbouring chain strands, typically \SIrange{1}{10}{\nano\meter} apart with a decay time of a few nanoseconds. On the other end of the spectrum, the slow mode has been reported to last milliseconds, and extend over a range of micrometers. It means that the correlation between the velocities of any two monomers can persist up to separations equal to thousands of chains apart.

Out of equilibrium, even a modest shear rate of $\kappa = \SI{1}{\second^{-1}}$ can easily couple to the slow mode, producing concentration correlations on roughly the same length scale as the quiescent velocity correlations. Since the size of these inhomogeneities is well above optical wavelengths, the otherwise clear solution turns opaque and turbid. Zooming in, the density fluctuations have been photographed under a microscope~\cite{mhetar1998slip}, showing ripples of \SI{10}{\micro\meter} and a characteristic ``butterfly pattern'' in the Fourier transform of the image. The time evolution of the butterfly has been elucidated using start up~\cite{van1992time, migler1996structure} as well as oscillatory~\cite{saito1999structures, saito2000light} shear flow. In the steady state, the scattered intensity has been measured to rise, plateau, and then a rise for a second time as a function of shear rate~\cite{hashimoto1992butterfly}. The second rise is associated with elastic turbulence, which just like inertial turbulence, is characterized by flow instabilities spanning the entire sample volume~\cite{groisman2000elastic, sousa2018purely}. The shape of the butterfly pattern has been accurately mapped over a broad range of distances using combined SALS and SANS data~\cite{saito2002structures}. In the flow-gradient plane $xy$, the butterfly was found to be tilted by \SI{40}{\degree} to the flow axis, and to rotate clockwise with increasing shear rate~\cite{wu1991enhanced}. This behavior is opposite to i.e. rods, which align at \SI{135}{\degree}.

The equilibrium fluctuations are well quantified by the Doi-Onuki theory~\cite{takenaka2007quantitative}, based on a time-dependent Ginzburg-Landau equation. However, the agreement under shear flow is more qualitative at this stage~\cite{takenaka2006computer}. An alternative approach by Helfand and Fredrickson was to couple the macroscopic equations for concentration, velocity, and stress fields~\cite{helfand1989large}. Their model was the first to explain the butterfly tilt angle, involving the balance of normal and shear forces on a concentration wave~\cite{ji1995concentration}. Milner has developed a two-fluid model, in the form of coupled Langevin equations~\cite{milner1993dynamical}. While instructive in their own right, up to now none of these theories could fit the experimental observations in a satisfactory manner~\cite{saito2001structure, saito2002structures}, although improvements are being made with recent papers examining extensional flow as well~\cite{cromer2013concentration, cromer2017concentration}.

Since the scale of concentration fluctuations is much greater than the size of one molecule, most theories have treated the system from a macroscopic continuum perspective. The equations tend to be rather complicated, including tensor fields with many variables and input parameters. No analytical solutions exist, so they have been solved numerically using various schemes like smoothed particle hydrodynamics~\cite{okuzono1997smoothed}. Given that computational work is inevitable, we may as well attempt a numerical simulation based on a particle model, which has an advantage of less assumptions and needs less theoretical insight. Entangled polymer solutions under shear have been simulated at several levels of coarse-graining~\cite{huang2010semidilute, korolkovas2018dynamical}, but nowhere near the scale required to probe the slow mode. The box would have to contain some $(\SI{10}{\micro\meter}/\SI{1}{\nano\meter})^3 = 10^{12}$ particles which is not an issue these days~\cite{potter2017pkdgrav3}. The real problem is the time needed to capture one fluctuation: $(\kappa \tau_e)^{-1} = 10^9$, where $\tau_e = \SI{1}{\nano\second}$ is the typical entanglement time. The longest trajectory so far reported for a hardcore model of Kremer-Grest type~\cite{grest2016communication} is $\SI{7e4}\tau_e$, and for a smaller softcore system of star polymers $\SI{9e6}\tau_e$ was achieved~\cite{korolkovas2018five}. At present, not even a glimpse of the slow mode is within reach of such models containing internal chain motion.

In this article we propose to coarse-grain the model to the level of the whole molecule, with one particle per chain. It can describe systems much larger than is feasible with bead-and-spring models, but has finer detail than the field theoretical approach. The intra-chain degrees of freedom are conveniently ignored, since it is known that the deformation of the form factor is minuscule~\cite{korolkovas2018anisotropy, Muller1993} compared to the $10^3$ increase of the structure factor intensity. Our main novelty is a method of reproducing inter-chain entanglement with a velocity-dependent friction force. It opposes any rapid change of inter-chain distance: overlapping chains resist separation, and separated chains resist being overlapped. Assuming that this friction has a linear dependence on the relative velocity (times a static envelope), the forces of the entire entangled system are solved exactly and reasonably fast, thanks to sparse matrix algebra. When shear is applied, large scale fluctuations emerge, shown in Fig.~\ref{box}.

\section{Method}
The goal behind this model is to make the concentration fluctuations emerge with the fewest possible assumptions, and with little computational power for ease of reproducibility and adaptation to more specific systems. At this stage the comparison to experiments remains only qualitative, but the merit of our simplified method is that all the structural changes can be traced back to essentially one interaction: velocity-dependent inter-chain friction. The chains are represented by undeformable particles, which we take for now to be spherical Gaussian clouds. Under shear, the chains are known to deform and become ellipsoidal, but this effect cannot be predicted from our coarse model, so it would have to be inserted by hand to better match experiment. Whichever particle shape is chosen, it is assumed to stay constant and not to depend on its neighbors. This is a reasonable proposition if the chains are not overlapping too much, meaning that the overlap density
\begin{equation}
\frac{\rho}{\rho^*} = \frac{3V}{4\pi C \lambda^3}
\end{equation}
is less than or equal to one (for strongest effect, we choose $\rho/\rho^* = 1$ in this demo). Here $C$ is the number of chains with radius of gyration $\lambda = \sqrt{Nb^2/6}$ in volume $V$, while $N$ is the number of monomers per chain and $b$ is the bond length. The pre-factor of $4\pi/3$ is only indicative, since the overlap $\rho/\rho^*$ can have somewhat different values depending on various theories and on the kind of experiment being considered, i.e. static (osmotic pressure) or dynamic (viscosity). Either way, the predictive power of our model is limited to situations where the polymers are entangled with not much more than their first shell of neighbors. Conveniently in this case, the SANS scattering cross section can be simplified to (Eq.~31.16 in Ref.~\cite{Hammouda2008}):
\begin{equation}\label{FS}
\frac{d\Sigma (\mathbf{q})}{d\Omega} = (\Delta \rho v N)^2 \frac{C}{V} \left( P(\mathbf{q}) + |F(\mathbf{q})|^2 [S(\mathbf{q})-1]\right)
\end{equation}
where $\Delta \rho$ is the scattering length density (SLD) contrast between polymer and solvent, and $v$ is the monomer volume. The average of the square form factor and the square of the average form factor are defined respectively by:
\begin{align}
P(\mathbf{q}) &= \braket{ \left|\sum_{n=1}^N \frac{e^{-i\mathbf{q}\cdot \mathbf{r}_n}}{N} \right|^2} = \frac{2\left( e^{-(q\lambda)^2} - 1 + (q\lambda)^2\right)}{(q\lambda)^4}\\
|F(\mathbf{q})|^2 &= \left|\sum_{n=1}^N  \frac{\braket{e^{-i\mathbf{q}\cdot \mathbf{r}_n}}}{N}  \right|^2 = \frac{\pi}{3}\left[\frac{\text{erf} (\sqrt{3}q\lambda/2)}{q\lambda} \right]^2 e^{-(q\lambda)^2/2}
\end{align}
The formulas on the right have been derived assuming an ideal random walk structure. At low $q\lambda \ll 1$ where the concentration fluctuations are strongest, both averages of the form factor are close to $P = |F|^2 = 1$, and the only relevant quantity is the inter-chain structure factor:
\begin{equation}\label{Sfactor}
S(\mathbf{q}) = \frac{1}{C}\braket{ \left|\sum_{c=1}^C e^{-i\mathbf{q}\cdot \mathbf{R}_c} \right|^2}
\end{equation}
We determine this quantity by direct simulation of the particles located at $\mathbf{R}_c = x_c \mathbf{\hat{x}} + y_c \mathbf{\hat{y}} + z_c \mathbf{\hat{z}}$. Without entanglements, the system is a soft colloidal suspension and is driven by the random, the excluded volume, and the shear forces:
\begin{subequations}
\begin{align}
\mathbf{F}_c &= \sqrt{6\zeta k_B T}\mathbf{W}_c(t)\\
&+ \frac{k_B T v}{\lambda^5} \sum_{c'=1}^C \mathbf{R}_{cc'} \exp \left(-\frac{\mathbf{R}_{cc'}^2}{2\lambda^2}\right)\label{exvol}\\
&+ \zeta \kappa(y_c-L_y/2) \mathbf{\hat{x}}
\end{align}
\end{subequations}
where $\braket{\mathbf{W}_c(t) \cdot \mathbf{W}_{c'}(t')} = \delta_{cc'}\delta(t-t')$ is the thermal noise modeled by a Wiener process. The friction coefficient $\zeta = 6\pi \eta_s Nb$ defines the natural unit of time $\tau = \lambda^2 \zeta/k_B T$ (the Rouse time). The temperature $k_B T$ is absorbed into the time unit and is not specified independently. Nevertheless, temperature plays a role in the thermodynamic quality of the (implicit) solvent, and is taken into account through the excluded volume parameter $v/\lambda^3$. An ideal solvent corresponds to $v/\lambda^3 = 1$, a borderline theta solvent is $v/\lambda^3 = 0$, while negative values are for a poor solvent. We have picked $v/\lambda^3 = 0.1$, which is still above the theta point, but not too high where the density fluctuations are suppressed. 

Our equation of motion is first order in time, meaning that the momentum is not conserved locally (the momentum dissipation time is much shorter than the coarse time step). This is in contrast to second order thermostats, such as dissipative particle dynamics (DPD), where the momentum is perfectly conserved, but we cannot use it because its time scale is much too short for our purpose. The lack of momentum conservation means that we have to impose the flow manually, and we choose a linear Couette profile of shear rate $\kappa$ in the $xy$ plane. Standard Lees-Edwards boundary conditions are used to simulate a bulk flow~\cite{korolkovas2018dynamical}. The velocity $\mathbf{v}_c = d\mathbf{R}_c/dt$ of each particle is obtained by coupling all the individual forces:
\begin{equation}\label{vc}
\mathbf{v}_c = \mathbf{F}_c/\zeta + \alpha \sum_{c'=1}^C (\mathbf{v}_{c'}-\mathbf{v}_c + \mathbf{V}_{cc'}) w_{cc'}
\end{equation}
The first term is the standard Brownian equation of motion, applicable in overdamped systems. The second term is our novelty to mimic the inter-chain entanglement force, whose physical motivation can be seen from the following thought experiment. Imagine two very long polymer chains in a big solvent bath, whose centers of mass (CM) can be displaced at will (by special optical tweezers). Initially the two chains are far apart and we bring them together so they interpenetrate. The speed of the CM is kept much smaller than the internal relaxation speed: $v_{cc'} \ll \lambda/\tau$ (which is an adiabatic, or a quasi-static process). The two chains have plenty of time to reptate around each other, resulting in zero elastic force throughout the process. The only force felt by the tweezers will be the excluded volume, Eq.~\eqref{exvol}. We repeat this experiment a second time, with the speed increased to $v_{cc'} \approx \lambda/\tau$ and the chains are brought from far away to right on top of each other $R_{cc'} = 0$. This is done quickly so the chains do not have time to reptate, resulting in an elastic repulsion, which to a first approximation is linearly proportional to the difference between the two velocities. For a very fast flow it may be necessary to include higher order terms (quadratic, cubic), but that is much more expensive to compute numerically, so for now we stick with the linear term. After some time, the chains will find a new equilibrium and the elastic force will subside to zero. We assume that the relaxation happens within the duration of one time step, which is reasonable if the speed is not excessively high. In terms of shear rate, $\kappa \tau \lesssim 1$. At higher speeds, the elastic force depends not only on the present velocity $v_{cc'}(t)$, but also on its recent history, and we do not consider this complication. Now the chains are entangled with each other, and we rapidly rip them apart. This time, an attractive force will be elastically generated, and is also assumed to be proportional to the speed $v_{cc'}$, which now has the opposite sign.

The equation of motion, Eq.~\eqref{vc}, describes all the situations of the thought experiment, and works simultaneously for a large system of particles. Note that the elastic force is non-conservative and cannot be derived from any Hamiltonian. It is fundamentally different from sticky attractive interactions which are prevalent in other highly viscous but unentangled fluids like bitumen. Topological entanglement, on the other hand, does not contain any enthalpy, and is only revealed in dynamical experiments. This is evidenced by the effect of the strength parameter $\alpha$, which strongly influences the dynamics (see Fig.~\ref{vv}), but leaves the static properties (structure factor, osmotic pressure) almost intact, as required. The entanglement range is assumed to be the same as the radius of gyration $\lambda$, and the coupling weight is chosen again to be Gaussian: $w_{cc'} = \exp\left(-\mathbf{R}_{cc'}^2/2/\lambda^2\right)$, but other reasonable shapes were seen to produce similar physics as well. A boundary correction term
\begin{equation}
\mathbf{V}_{cc'} = \kappa L_y \texttt{round}(y_{cc'}/L_y) \mathbf{\hat{x}}
\end{equation}
is necessary in case the particle separation $y_{cc'}$ is larger than half of the box height $L_y$. To solve Eq.~\eqref{vc}, we rewrite it in the standard linear algebra form:
\begin{equation}\label{algebra}
\begin{pmatrix}
1+\sum w_{1c} & -w_{12} & -w_{13} & \cdots\\
-w_{21} & 1+\sum w_{2c} & -w_{23} & \cdots\\
\vdots & & \ddots
\end{pmatrix} \begin{pmatrix}
\mathbf{v}_1\\ \mathbf{v}_2\\ \vdots
\end{pmatrix} = \begin{pmatrix}
\mathbf{F}_1 + \sum \mathbf{V}_{1c} w_{1c}\\
\mathbf{F}_2 + \sum \mathbf{V}_{2c} w_{2c}\\
\vdots
\end{pmatrix}
\end{equation}
We truncate the weights below a chosen threshold (\SI{5}{\percent}), and use a sparse matrix solver to boost the code performance. The MATLAB source code is available as a Supplementary material.

\section{Results}
\begin{figure}[ptbh!] 
		\begingroup
			\sbox0{\includegraphics{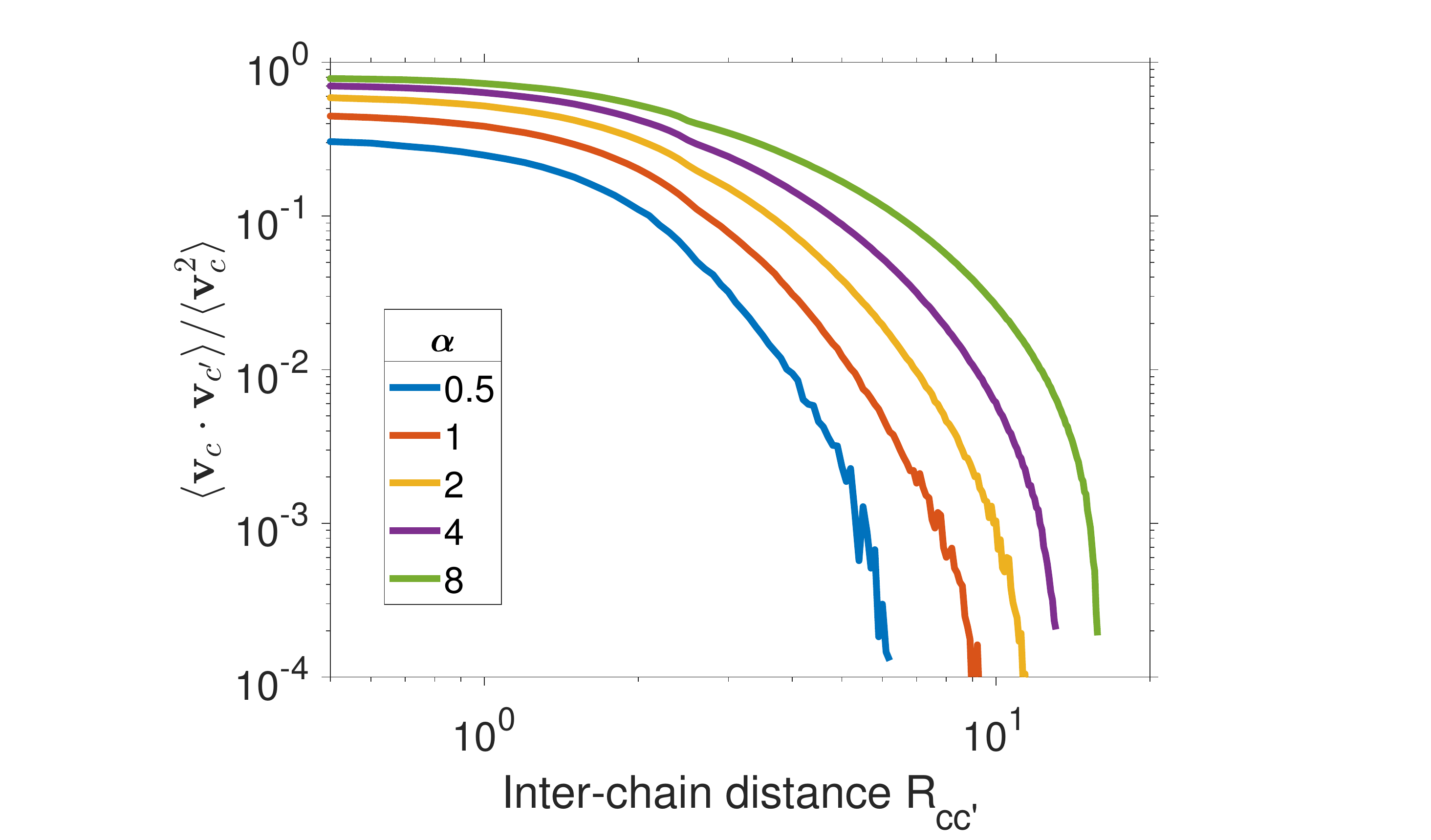}} 
			\includegraphics[clip,trim={.1\wd0} 0 {.15\wd0} 0,width=0.6\linewidth]{./fig/vv.pdf} 
		\endgroup
\caption{The average inter-chain velocity product as a function of inter-chain distance, for various entanglement strength values $\alpha$. }\label{vv}
\end{figure}

While our method cannot resolve overlapping particles $\rho/\rho^*\gg 1$, it provides a robust solution for practically any value of the entanglement coupling strength $\alpha$. Since Eq.~\eqref{algebra} is solved exactly with near perfect convergence, we can model scenarios ranging from small $\alpha\ll 1$ (non-entangled soft colloidal suspension), to huge $\alpha \gg 1$ where the entire box is coupled into one solid body, and could be used to model a (transient) gel. The effect of $\alpha$ is shown in Fig.~\ref{vv}, where the velocity cross-correlation $\braket{\mathbf{v}_c\cdot \mathbf{v}_{c'}}/\braket{\mathbf{v}_c^2}$ is seen to persist for ever longer distances $|\mathbf{R}_{cc'}|$, as expected from DLS measurements~\cite{li2010slow}. 

\begin{figure}[tbhp!] 
			\includegraphics[width=0.49\linewidth]{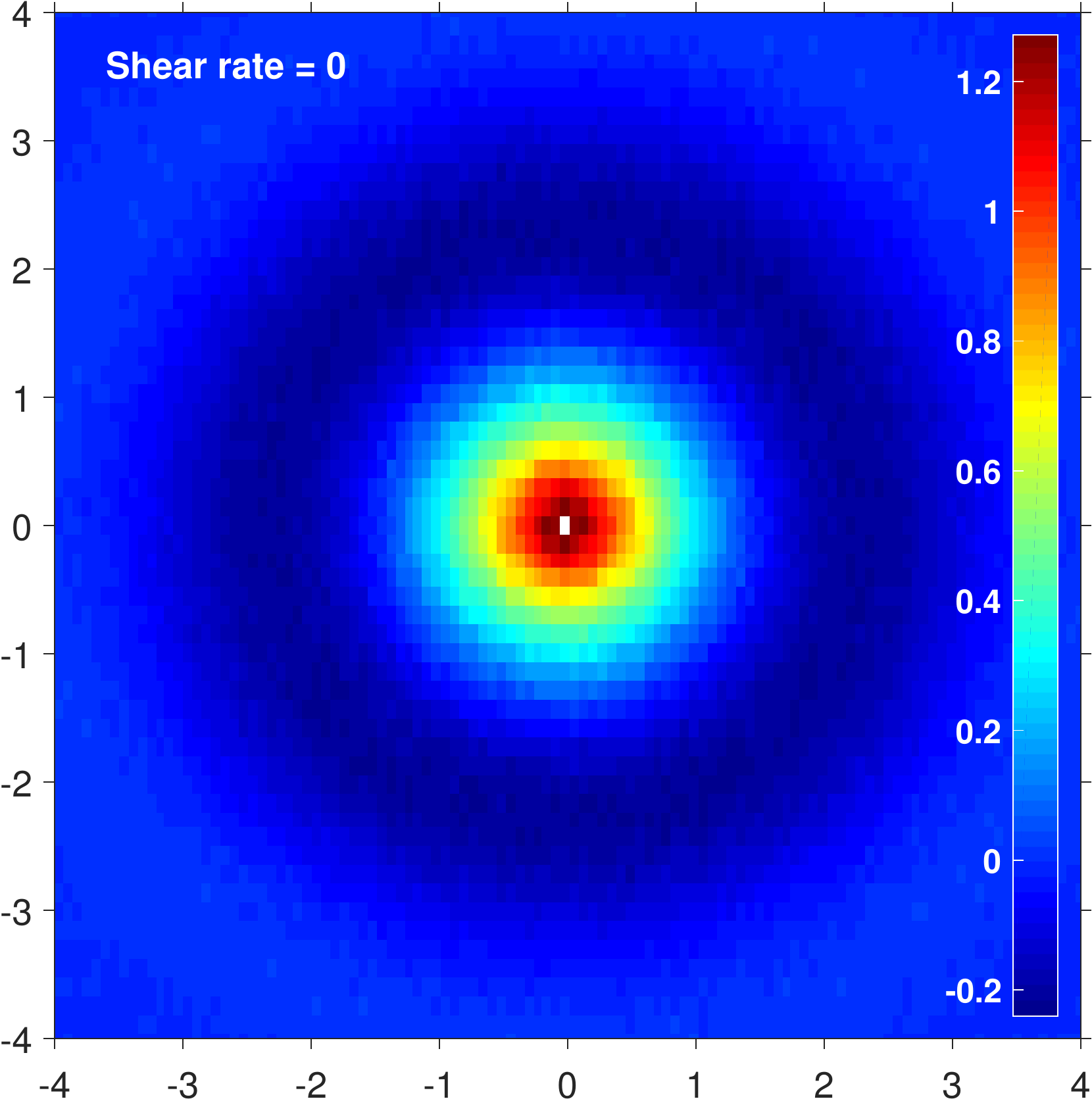}\hfill
			\includegraphics[width=0.49\linewidth]{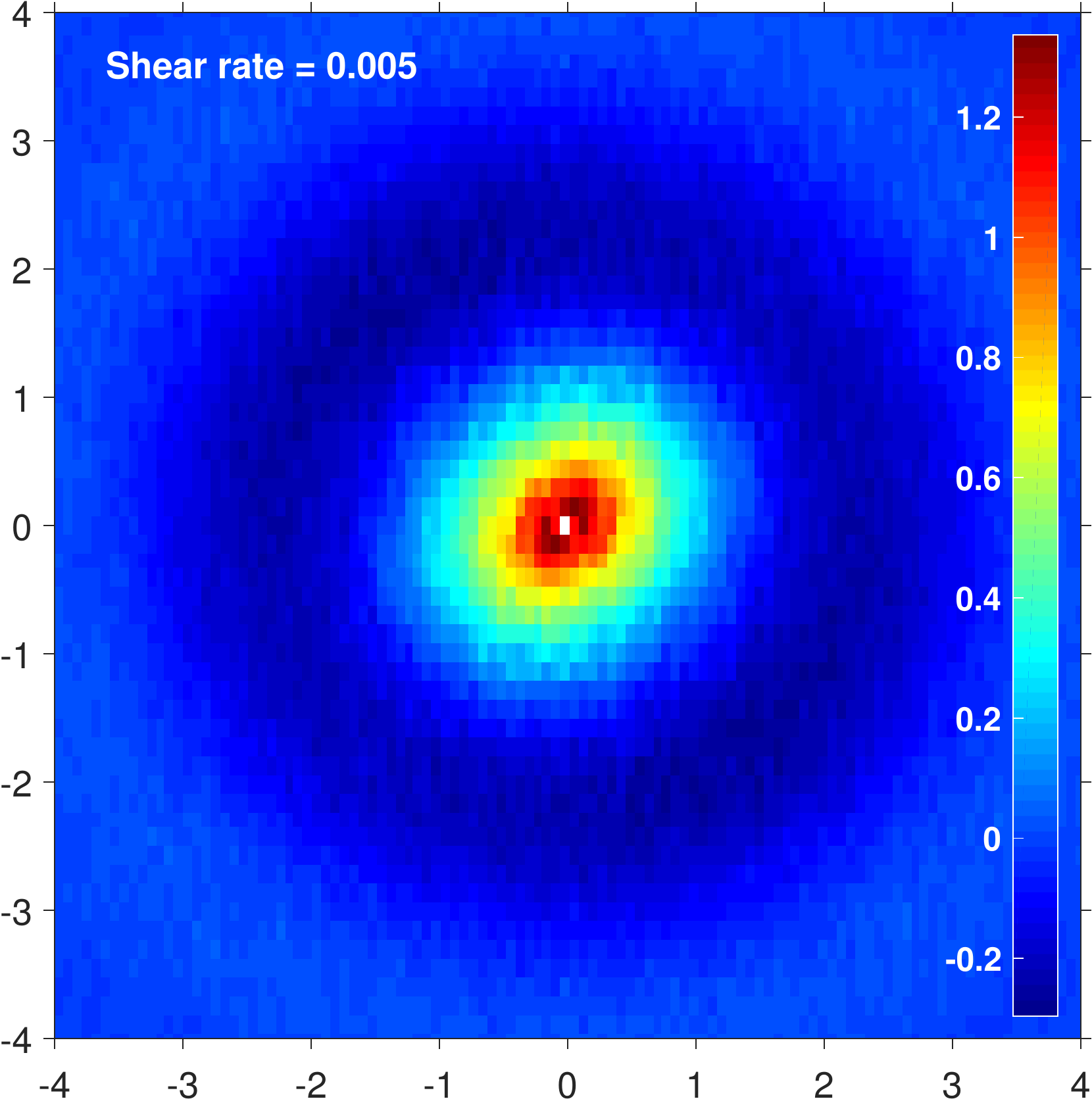}
			\includegraphics[width=0.49\linewidth]{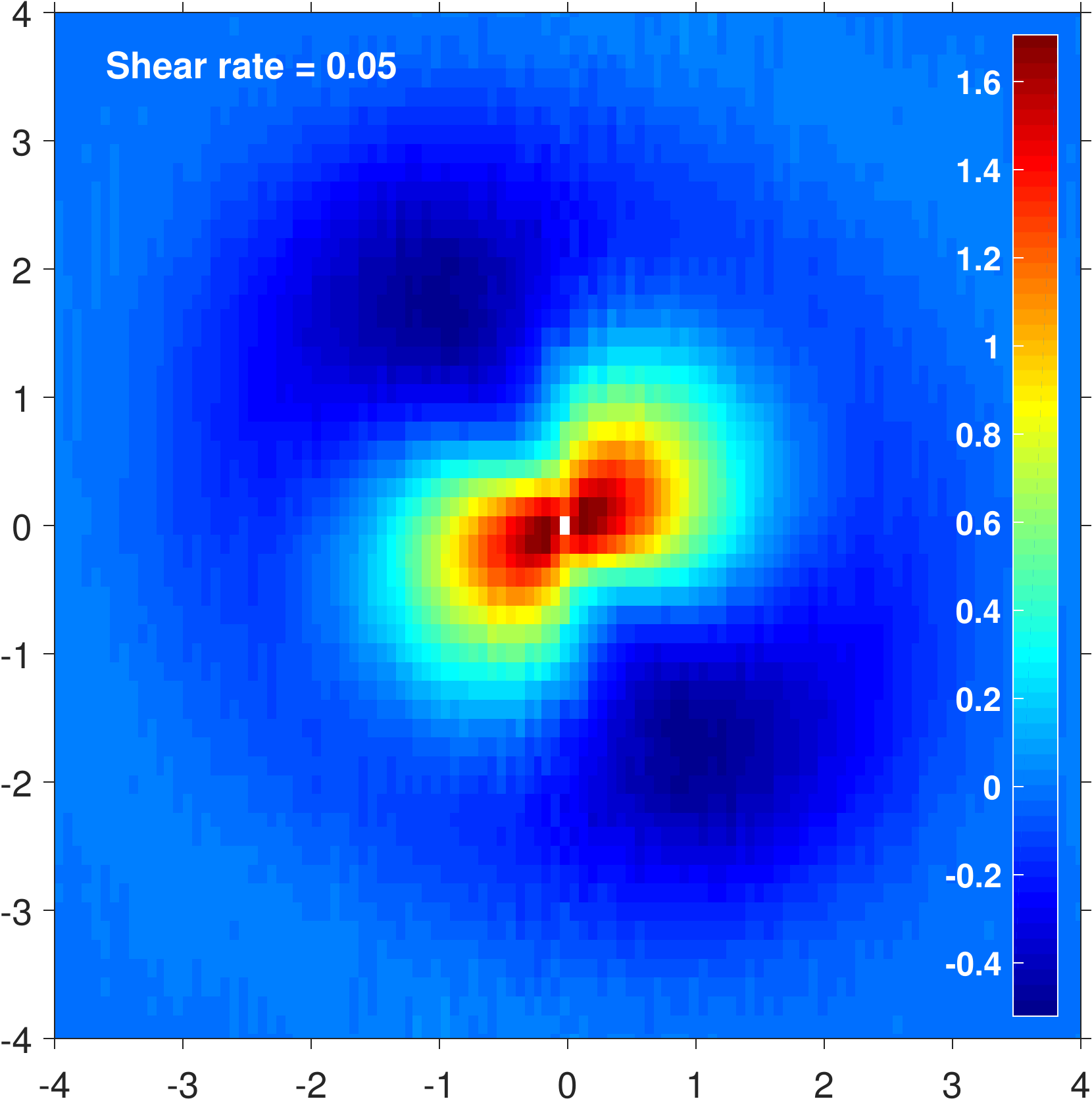}\hfill
			\includegraphics[width=0.49\linewidth]{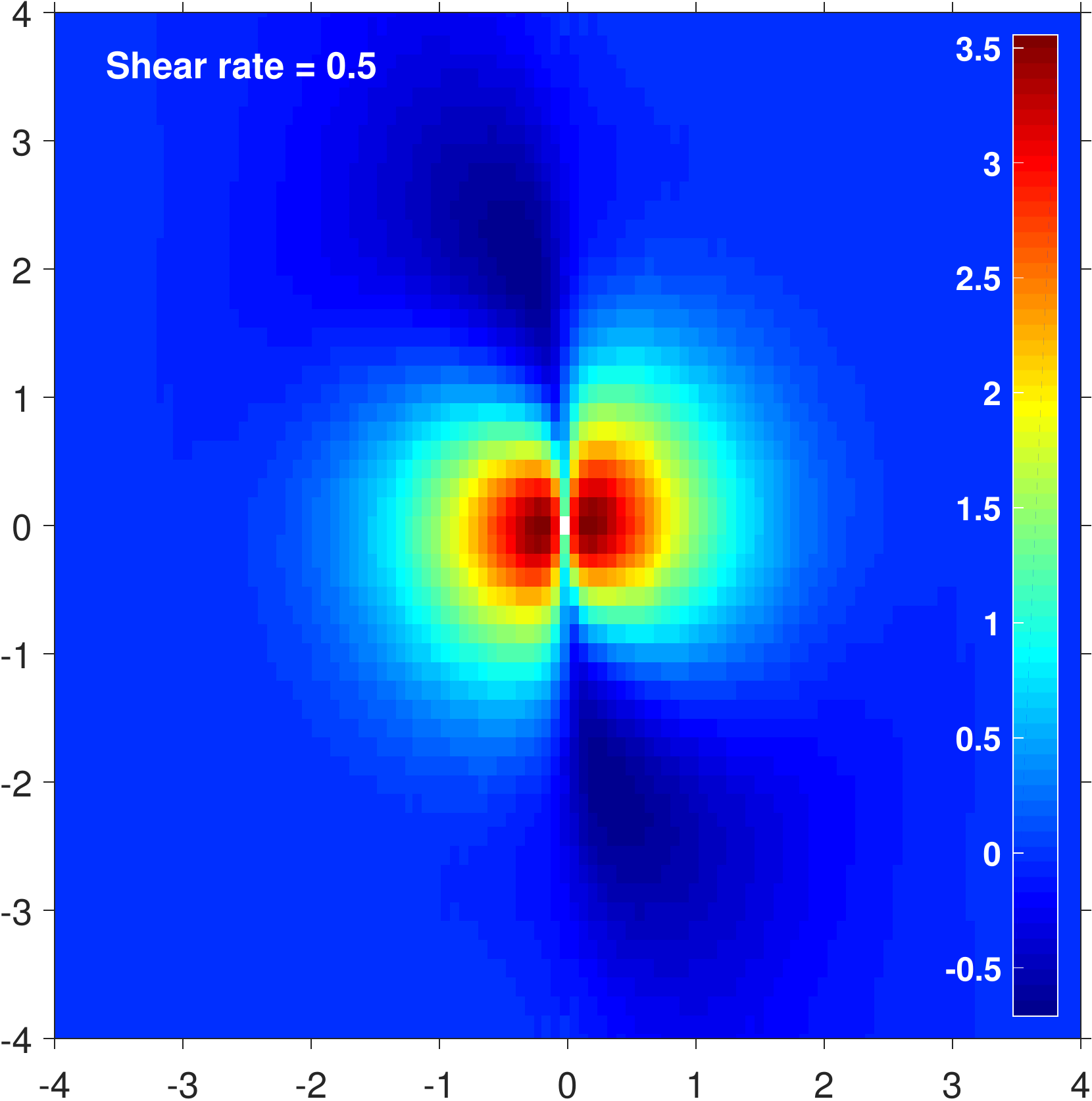}
\caption{The log of the structure factor (Eq.~\eqref{Sfactor}) along the flow ($q_x \lambda$, horizontal) and shear gradient ($q_y \lambda$, vertical) directions. The shear rate is given in dimensionless units $\kappa \tau$. Notice the increase of intensity $e^{3.5/1.2}\approx 20$ between zero and maximum shear.}\label{Sxy}
\end{figure}

For the shear simulation we have chosen $\alpha=4$, where the correlation length is about $|\mathbf{R}_{cc'}| = 13\lambda$. A rectangular box of volume $V = 87.6 \times 43.6\times 22 \lambda^3$ was used, with the smallest side sufficiently wide to avoid finite size effects. Four different shear rates were applied, and the resulting inter-chain structure factors are shown in Fig.~\ref{Sxy}. (The initial startup trajectory was discarded, so the data shows steady-state structure). At zero shear, we see a typical soft liquid structure factor with a minimum at $q\lambda = 2.3$, which corresponds to the first shell of neighbors. At shear $\kappa \tau = 0.005$, this structure is skewed along the flow, meaning that the pair correlation function is now skewed backwards: there are more neighbors in quadrants 2 and 4, than 1 and 3. We can already see a small peak appearing at $q_x\lambda = 0.1$, which corresponds to the distance of correlated velocities in Fig.~\ref{vv}. Increasing the shear to $\kappa \tau = 0.05$, the peak grows in intensity, creating a peanut-shaped pattern. At highest shear $\kappa \tau = 0.5$, the intensity grows further and the pattern rotates towards $q_y=0$ axis, assuming a butterfly pattern. In real space, the highest shear run is shown in the Supplementary Video, from which a snapshot was plotted in Fig.~\ref{box}.

\begin{figure}[tbh!]
		\includegraphics[width=0.6\linewidth]{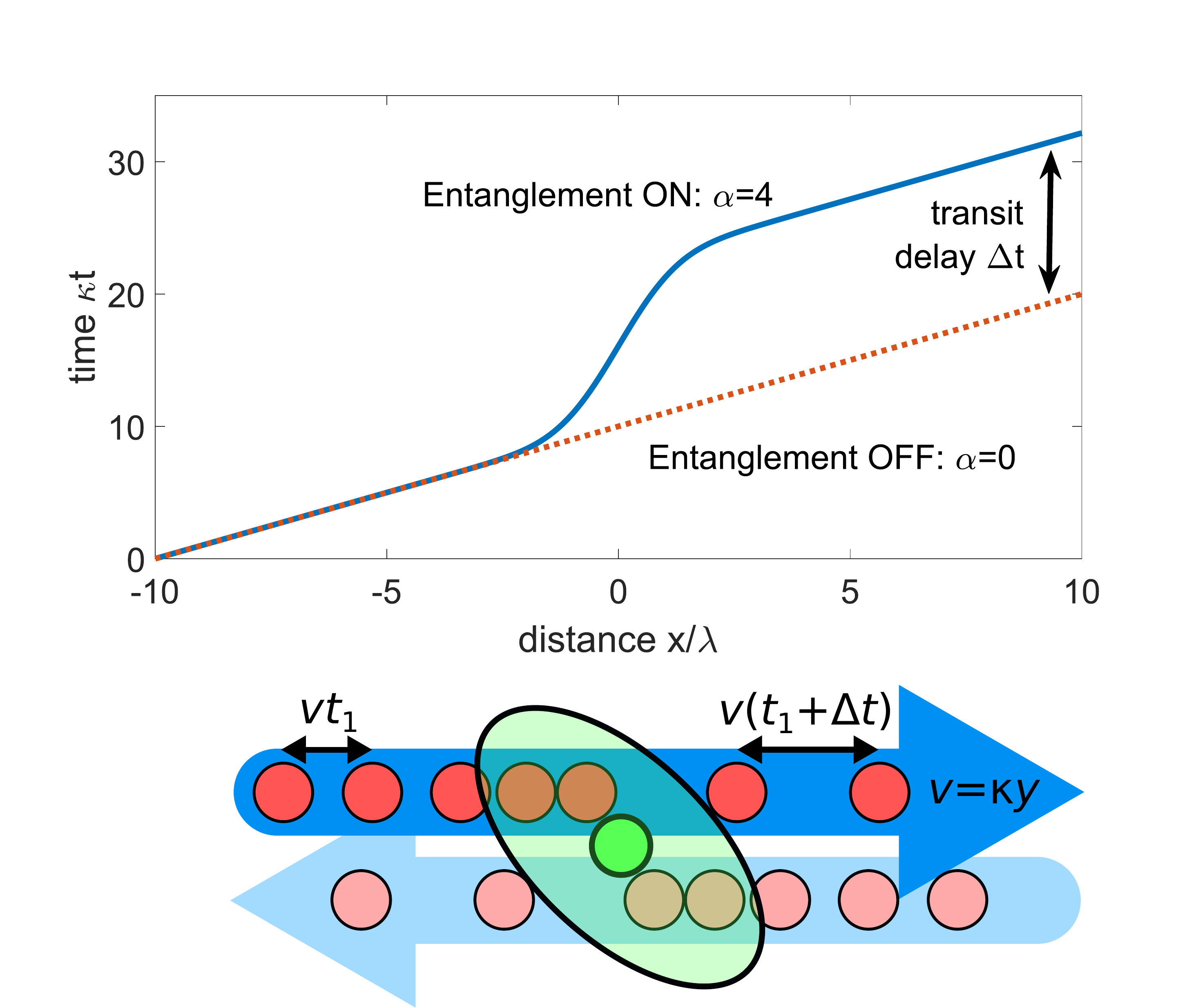}
\caption{The inception of a concentration fluctuation. Incoming red particles are slowed down by the obstacle, resulting in a time delay $\Delta t$ (upper panel, Eq.~\eqref{timedelay}). A dense stream of particles (lower panel) is partially blocked, forming a traffic jam (the green-tinted oval, aligned backwards to the flow).}\label{trafficjam}
\end{figure}

To pinpoint the crux of the fluctuation phenomenon, let us consider just two particles at a height $y$ apart, and ignore for now the excluded volume as well random forces. From Eq.~\eqref{algebra} we obtain the velocity $dx/dt = v/(2w+1)$, where $w = \alpha e^{-(x^2+y^2)/2/\lambda^2}$ and $v=\pm \kappa y$ is a constant speed, with opposite signs for the two particles. The integral is
\begin{equation}\label{timedelay}
(x_2-x_1)/\lambda + \sqrt{2\pi}\alpha e^{-(y/\lambda)^2/2} [\text{erf}(x_2/\lambda/\sqrt{2})-\text{erf}(x_1/\lambda/\sqrt{2})] = \kappa y t/\lambda
\end{equation}
Assuming $y=\lambda$, we plot this equation in Fig.~\ref{trafficjam} (upper panel). The entangled case with $\alpha=4$ takes a time $\Delta t$ longer to pass by the obstacle, but notice that the elastic repulsion on the incoming side and the attraction on the outgoing side are of equal magnitude, resulting in a symmetric structure. In the lower panel we consider not just two, but a stream of incoming particles. If the delay $\Delta t$ is greater than the spacing $t_1$, there will be a traffic jam behind the obstacle. Eventually a steady state is reached, when the pressure of the traffic jam grows to the size where the number of incoming and outgoing particles is equal.

\section{Discussion}
Under a fast shear flow, the entangled chains do not have enough time to follow their rapidly flowing neighbors, resulting in deformations both on the scale of an individual chain (the form factor), and the density of the system as a whole (the structure factor). The osmotic pressure in semi-dilute solutions can be quite low, favoring an increase of density inhomogeneities which reduce the friction between the flowing chains, thereby mitigating their internal deformation. In this simulation, we have completely disregarded the form factor, and isolated the role of just the structure factor. While this is sufficient to induce large concentration fluctuations, certain details do not match experimental facts. In this section we discuss the limitations of our minimalist method, and how the consideration of the form factor may reconcile it with experiment. Other theories introduce phenomenological functions like Rolie-Poly to better match experiment~\cite{cromer2017concentration}, but that amounts to adding an assumption to the model, not extracting a prediction from it, and we do not pursue that strategy.

\textbf{The scattering intensity.} Experimentally, the scattering intensity has been measured to increase by over a factor of $10^3$ in Ref.~\cite{saito2002structures}. The increase seen in our simulation is merely a factor of 20, and is principally limited by the size of the box that was convenient to work with on a desktop PC. On a supercomputer, much bigger systems can be envisioned, enabling a greater $\alpha$ (see Fig.~\ref{vv}), hence stronger density fluctuations. The intensity also increases with shear rate, and our model is expected to be reasonably realistic for $\kappa \tau<1$. Beyond that, the internal chain dynamics may start to play a crucial role, and non-linear terms may become necessary in Eq.~\eqref{vc}. Further, for strong deformations the velocity of the particle may depend not only on the present state, but also on the history of the nearby trajectories. Lastly, in the regime of elastic turbulence~\cite{groisman2000elastic, sousa2018purely} we need to consider truly macroscopic dimensions, on the scale of the whole shear apparatus, in which case a particle simulation is too costly even on a supercomputer, and a coarser continuum model is warranted.

\textbf{The position of the peak.} The hallmark of shear induced concentration fluctuations is the peak in intensity at small but finite $q$ along the flow axis $x$. Experiments report the shift of this peak towards lower $q$ with increasing shear. The difference between the two peaks visible in Ref.~\cite{saito2002structures} is about \SI{50}{\percent}, which is quite small (compared to the tenfold increase of intensity for the same curves), and coincides with the extent of a typical form factor deformation~\cite{Muller1993}. In our simulation, the particles are assumed to be undeformable, resulting in a peak which does not budge across the full range of shear rates. For us, the sole option to make the peak move is to use phenomenologically elongated particles, which would shift the peak to a proportionally lower $q$. Equivalently, if the experimental data would be rescaled to the measured radius of gyration at each given shear rate~\cite{korolkovas2018anisotropy}, we expect the peak to stay at a fairly constant dimensionless $q\lambda(\kappa)$.

\textbf{The rotation of the peak.} The form factor of sheared polymers is aligned at \SI{135}{\degree} to the flow axis, and this angle decreases with the shear rate~\cite{Muller1993}. Similar alignment is seen in most other anisotropic systems, i.e. carbon nanotubes~\cite{fan2005characterization}. In contrast, the structure factor of shear induced concentration fluctuations orients at \SI{45}{\degree}, and rotates clockwise with higher shear, see Fig.~\ref{Sxy}. In our simulation, the angle approached close to \SI{0}{\degree} at highest shear, where it saturated. Experimentally, however, the peak is seen to cross over below \SI{0}{\degree} under sufficient shear~\cite{wu1991enhanced, van1992time}. In field theories, the peak rotation past the flow axis has only been recently been demonstrated~\cite{cromer2017concentration}, by incorporating the Rolie-Poly constitutive equation to describe polymer stress. By analogy, if we were to replace our spherical particles with phenomenological ellipsoids, the fluctuations will be reshaped accordingly, resulting in the scattering peak below the flow axis. In addition, the scattering cross-section is the product of the form and the inter-chain structure factors, see Eq.~\eqref{FS}, so multiplying our $S(\mathbf{q})$ data with a form factor envelope $|F(\mathbf{q})|^2$ of a tilted ellipsoid, one can obtain the rotation of the peak past the flow axis as well.

\textbf{The dark streak.} The plane defined by $q_x=0$ is known as the ``dark streak'' since its scattering is much weaker than elsewhere. Nevertheless, experiments report an increase of scattering in all directions, including the dark streak, although it always stays weaker than the one along the flow axis~\cite{saito2002structures}. By contrast, the intensity of the central pixel $q_x = 0$ in our simulation remains almost the same regardless of shear rate. Given that the structure factor is the Fourier transform of the pair correlation function~\cite{squires2012introduction}, and the latter is the probability density whose integral is fixed, we find it strange that the scattering increases everywhere. We speculate that it may be an experimental artifact in both Couette and parallel plate geometries, where neutrons probe slightly curved flow fields. While the neutron momentum transfer along $q_y$ contains mostly scattering along the shear gradient $y$, a small percentage from the flow axis $x$ contributes to $q_y$ as well. If the intensity along the flow increases by $10^3$ of which \SI{1}{\percent} is registered along the $q_y$ axis due to the sample curvature, an increase of 10 will be erroneously reported. This controversy could be settled by a future experiment with a narrow beam footprint to minimize the cross-contamination between the curved axes, as well as high $q$ resolution along the flow to distinguish the tiny central pixel from the massive intensity peaks on both of its sides.

\begin{acknowledgement}
The author acknowledges financial support of the Swedish research council and Carl Tryggers stiftelse, grant CTS 16:519.
\end{acknowledgement}



\bibliography{manuscript}

\end{document}